\documentclass[a4paper]{article}

\usepackage{INTERSPEECH2022}

\title{Adversarial Multi-Task Deep Learning for Noise-Robust Voice Activity Detection with Low Algorithmic Delay}
\name{Claus M. Larsen, Peter Koch, Zheng-Hua Tan}
\address{
Department of Electronic Systems, Aalborg University, Denmark }
\email{zt@es.aau.dk}

\begin{document}

\maketitle
\begin{abstract}
Voice Activity Detection (VAD) is an important pre-processing step in a wide variety of speech processing systems. VAD should in a practical application be able to detect speech in both noisy and noise-free environments, while not introducing significant latency. In this work we propose using an adversarial multi-task learning method when training a supervised VAD. The method has been applied to the state-of-the-art VAD \textit{Waveform-based Voice Activity Detection}.  Additionally the performance of the VAD is investigated under different algorithmic delays, which is an important factor in latency. Introducing adversarial multi-task learning to the model is observed to increase performance in terms of Area Under Curve (AUC), particularly in noisy environments, while the performance is not degraded at higher SNR levels. The adversarial multi-task learning is only applied in the training phase and thus introduces no additional cost in testing. Furthermore the correlation between performance and algorithmic delays is investigated, and it is observed that the VAD performance degradation is only moderate when lowering the algorithmic delay from 398 ms to 23 ms.

\end{abstract}

\section{Introduction}

Voice Activity Detection (VAD) aims to detect which segments of an audio stream contains speech and the segments are typically 10 ms each \cite{rVAD}, \cite{preproc1}, \cite{WVAD}. It is widely used as a pre-processing step in more complex audio signal processing tasks such as speech recognition \cite{baevski2021unsupervised}, speaker verification \cite{chettri2020dataset} or speech enhancement \cite{hoang2022minimum}, but can also be applied on its own to reduce the computational cost of downstream processing. While detecting speech in a noise-free environment is a trivial task, the difficulty in classification arises in noisy environments \cite{dinkel2021voice}.

VAD algorithms can generally be categorized into classes of supervised and unsupervised methods. The unsupervised methods can be energy based \cite{energybased_1}, however, this approach is very sensitive to noisy conditions. More complex unsupervised methods are generally based on assumptions of speech and noise characteristics, e.g. using Mel Frequency Cepstral Coefficients (MFCCs) \cite{VQVAD}, \cite{MFCC}, perceptural spectral flux \cite{sadjadi2013unsupervised} or pitch detection and spectral flatness \cite{rVAD}.

In recent years supervised methods for VAD has gained increased popularity within the field of research \cite{rho2022vad}, \cite{lee2019spectro}. The supervised methods require large amounts of labelled speech data and their performance are highly dependent on the quality of the labelled data used for training and testing. Some supervised methods contains a pre-processing step which aims to extract useful features from the audio such as MFCCs \cite{zhang2013deep}, while other methods resolves to the raw waveform as their input \cite{WVAD}, \cite{CLDNN}. 
A benefit of using the raw waveform as input to the supervised VAD is that the method will potentially find the most optimal features to be used for classification on its own, and is therefore able to utilise both the magnitude and the phase of the audio \cite{WVAD}. Using the raw waveform as features to the VAD is an active area of research and has shown appealing performance in terms of noise-robustness \cite{WVAD}, \cite{CLDNN}. 

Two important factors in a VAD algorithm are how noise-robust it is and how much latency it introduces. Adversarial multi-task learning has proven to be effective to make applications invariant to noise and thereby more noise robust, e.g. for speech recognition in \cite{multi_task} and speech enhancement in \cite{mengadversarial}. Noise robustness of speaker verification is largely boosted by adversarial training in \cite{yu2017adversarial}. 
When applying VAD in a real-world application, often the latency is of great concern. Even though VAD is an active area of research, the algorithmic delay it introduces is rarely explored. 

In this work we aim to investigate if the noise-robustness can be increased even further by introducing an additional sub-network which aims to classify the noise types to a supervised VAD, and train the VAD adversarially to these. The supervised VAD method presented in \cite{WVAD}, which is based on fully convolutional neural networks (CNN) and shows state-of-the-art performance on the AURORA2 dataset \cite{AURORA}, will be used as the framework in this work. An additional discriminative sub-network for adversarial multi-task training, inspired by the work of \cite{DOA_original} and further refined for the speech recognition task in \cite{multi_task} will be introduced. The additional network used for adversarial multi-task learning is only introduced in the training phase and thus introduces no additional cost in testing. Furthermore, we investigate the impact of different algorithmic delays on VAD performance and realise this by varying CNN kernel sizes.  
The source code for this work is publicly available on GitHub \footnote{https://github.com/aau-es-ml/VAD-with-adversarial-multi-task-learning}.


\begin{figure*}[!htp]
    \centering
    \includegraphics[width=0.8\textwidth]{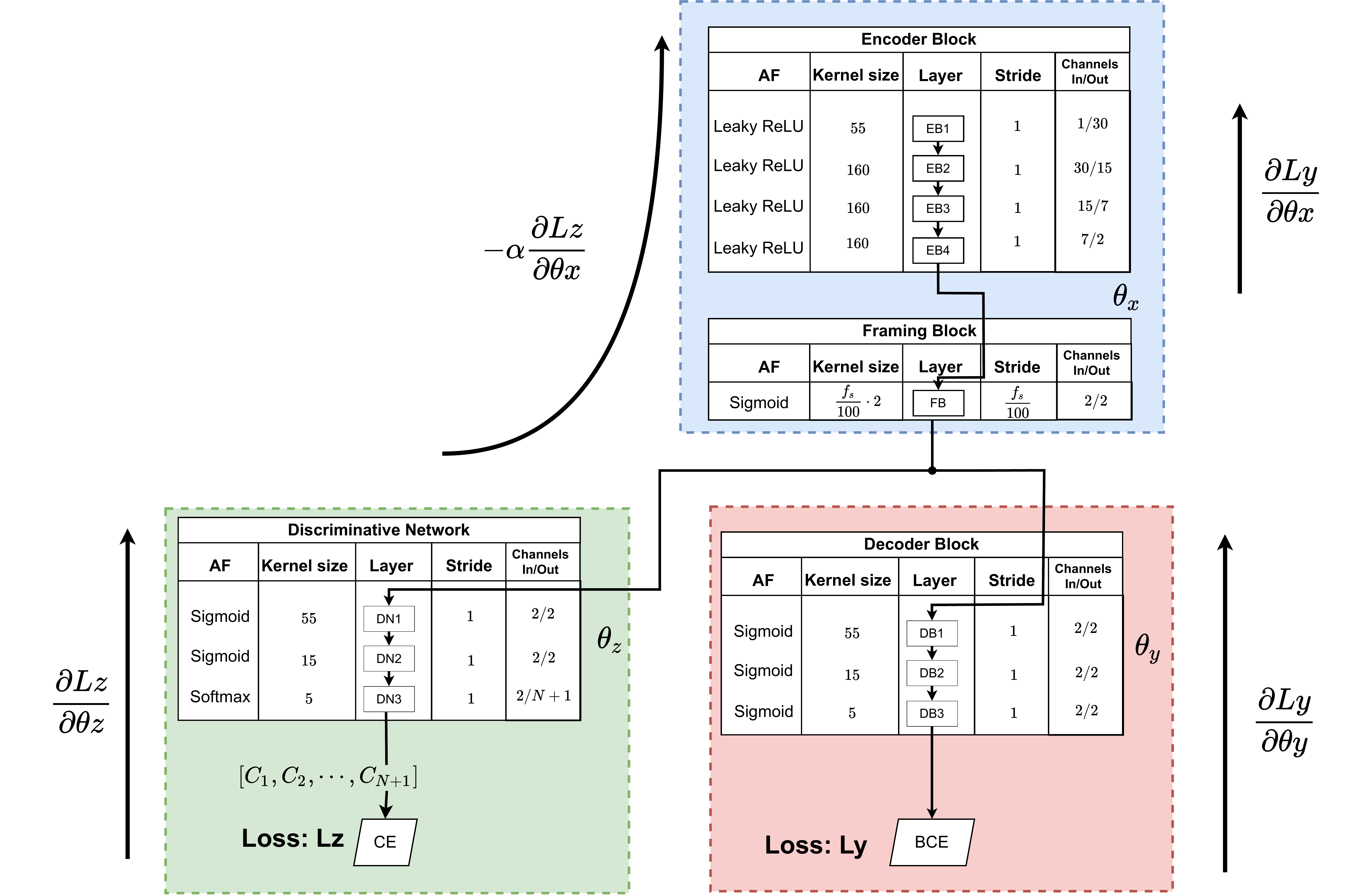}
    \caption{Overview of the proposed discriminative network (the green block) applied to the waveform-based VAD (the blue and red blocks) proposed in \cite{WVAD}.}
    \label{fig:proposed_method}
\end{figure*}

\section{Proposed method}
In this work we propose to introduce adversarial multi-task learning to enhance the robustness of deep model based VAD. Specifically, an additional sub-network for adversarial training is introduced to a state-of-the-art waveform-based VAD using a CNN model. 
The entire framework is illustrated in Figure \ref{fig:proposed_method}, in which the adversarial-training sub-network is shown by the green box, the algorithmic delay is investigated by modifying the blue block, and the waveform-based VAD \cite{WVAD} consists of the blue and red blocks.

\subsection{Framework}

In realising the adversarial multi-task learning for VAD, we consider the model presented in \cite{WVAD} with our own implementation in Pytorch \cite{PyTorch}. 
The method resorts to a fully convolutional neural network. It consists of an 
Encoder Block (EB), a Framing Block (FB) and a Decoder Block (DB). 
The EB is taking as input the raw waveform of audio and the FB generates features on a 10 ms basis. The DB further refines these features into the VAD output, where the larger value of the two channels, \textit{speech and non-speech}, determines the VAD label as speech or non-speech on a 10 ms basis.

In this work we introduce the additional Discriminative Network (DN). It takes as input the output from the FB and aims to learn to correctly classify the different noise types as well as to adversarially train the EB and FB part to make it noise robust.

\subsection{Adversarial multi-task learning}

The first part of this work is to introduce adversarial learning. The proposed DN is implemented in a similar way to the DB, where each layer resolves time-frequency representations in the channels of feature maps and the shrinking kernel sizes (55, 15, 5) reflect the decreasing modulation frequency (1.83 Hz, 6.66 Hz, 20 Hz) with the segment rate being 100 Hz \cite{WVAD}. The discriminative network generates softmax probabilities for each of the $N+1$ channels (i.e. outputs), where $N$ is the number of different noise types in the training set, while the remaining channel is for clean speech. The labels used for the DN is the noise types on a 10 ms basis, similarly to the VAD labels.

Following the DN, the cross-entropy loss is calculated between the noise types predicted by the DN and the true noise types. Following the DB is the binary-cross-entropy loss calculated between the VAD labels and the truth labels.
The losses are expressed as:

\begin{align}
    L_{z} &= -\sum_{i} t_i \mathrm{log}(p_i) \\
    L_{y} &= - \left[t \mathrm{log}(p)+(1-t) \mathrm{log}(1-p) \right] 
\end{align}
where \textit{t} is the true labels and \textit{p} is the scores output by the networks on a 10 ms basis.

When backpropagating the error through the model, the gradients of the DN are updated based only on the loss $Lz$, the gradients of the DB are updated based only on the loss $Ly$ while the gradients of the EB and the FB are updated based on both losses. However, the sign of the gradients calculated from $Lz$ is flipped such that the EB and FB are trained adversarially to the DN and friendly to the DB. Additionally the magnitude of this gradient is multiplied by a scalar $\alpha$ that determines the contribution from this sub-network. The key idea behind the method is that the FB will then output features that are invariant to the noise type which in turn will lead to a more noise-robust VAD and hence better VAD performance. The DN is illustrated in Figure \ref{fig:proposed_method}. The gradients are noted as the partial derivatives of the loss function $L$ with respect to the parameters $\theta$.

The convolutional operations of the DN can be expressed as:


\begin{equation}
    y_{[c]}^{[l]}(\tau) = \mathrm{AF}\left(\left(\mathbf{F}_{[c]}^{[l]} * y_{[c]}^{[l-1]}(\tau)\right) + \mathbf{b}_{[c]}^{[l]} \right)
\end{equation}
where $c$ denotes the channel, $l$ denotes the layer, $\mathbf{F}_{[c]}^{[l]} \in \mathbb{R}^{C \times k}$ is the convolutional kernel, $\mathbf{b}_{[c]}^{[l]} \in \mathbb{R}^{C \times 1}$ is the bias, $y_{[c]}^{[l]} \in \mathbb{R}^{C \times \mathrm{max}(\tau)}$ is the feature map and AF is the activation function.

\subsection{Algorithmic delay}
\begin{figure}[!htp]
    \centering
    \includegraphics[width=0.48\textwidth]{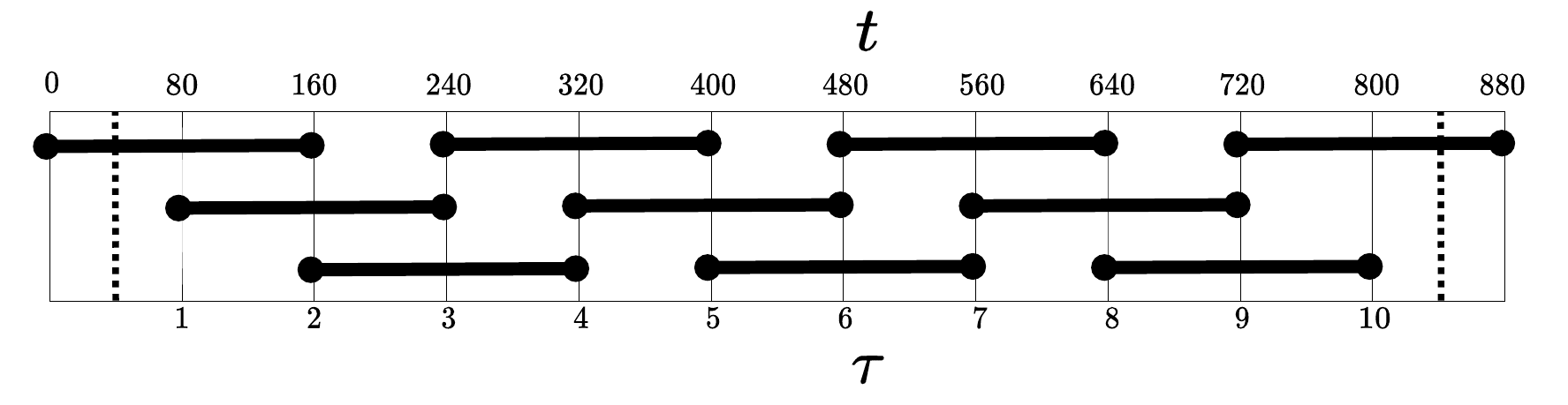}
    \caption{Illustration of the context needed to generate an output from the FB. In this example 10 outputs are generated from an 8 kHz signal, but because of the 50\% overlap between frames, an additional ${\frac{fs}{100}}-1$ samples of context is needed to generate an output.}
    \label{fig:framing_block}
\end{figure}

Second part of this work focuses on reducing the latency of the VAD. Only the algorithmic delay is considered, i.e. it is investigated how the VAD performance is affected based on how many future samples are used in the predictions, from here on referred to as \textit{future context}. The amount of future temporal context used for a given classification is calculated based on the fact that the feature map through a 1-dimensional convolutional layer will shrink as stated by Eq. \ref{eq:conv_shrink}.

\begin{equation}
    n^{[l]}-k+1 = n^{[l+1]}
    \label{eq:conv_shrink}
\end{equation}
where $n^{[l]}$ is the size of the feature map generated by the $l^{th}$ layer and $k$ is the kernel size.
The number of samples by which the feature map is shrinking will have to be considered as context where half of it is past and the other half is future.
The algorithmic delay introduced by the network is found by calculating how much context is needed in each layer and finally summing them together. The context introduced in the EB layers is simply found as in Eq. (\ref{eq:conv_shrink}), while the context introduced in the framing block is more complex and best illustrated by Figure \ref{fig:framing_block}. The context introduced by the DB is dependent on the stride of the FB and once again calculated using Eq. (\ref{eq:conv_shrink}). The total algorithmic delay in seconds is found by dividing the context with two times the sampling rate and is calculated as Eq. (\ref{eq:context}).

\begin{equation}
   AD = \frac{\sum_{n=1}^{4} (k_{EBn}-1) + \frac{fs}{100}-1 + \sum_{i=1}^{3} (k_{DBi}-1) \cdot \frac{f_s}{100}}{2f_s}
    \label{eq:context}
\end{equation}

\section{Speech corpora}
In this work two speech corpora are used. The AURORA2 \cite{AURORA} database and the TIMIT \cite{TIMIT} database.
\subsection{AURORA2}

 First is the AURORA2 \cite{AURORA} database which is used for multi-condition training at a sampling frequency of 8 kHz. The training set consists of 8440 utterances with four noise types artificially added at SNR levels of  $5 \ \text{dB}$, $10 \ \text{dB}$, $15 \ \text{dB}$, $20 \ \text{dB}$ and \textit{clean}. The four noise types used are \textit{subway, babble, car} and \textit{exhibition hall}. For each combination of noise type and SNR level 422 utterances are used.

AURORA2 contains three test sets, of which two are used in this work. Test set A uses the same noise types as the training set, and test set B uses four noise types unknown to the training set. These are \textit{restuarant, street, airport} and \textit{train station}. In the test sets the following SNR levels are used: $-5 \  \text{dB}$, $0 \ \text{dB}$, $5 \  \text{dB}$, $10 \ \text{dB}$, $15 \ \text{dB}$, $20 \ \text{dB}$ and \textit{clean}. Each test set consists of 4004 utterances which are evenly distributed on the four noise types and repeated for each SNR level. The true VAD labels are from the open-source rVAD repository \cite{rVAD}.

\subsection{TIMIT}

Secondly, the TIMIT \cite{TIMIT} speech corpus is used. The training set and test set, respectively, consists of 4620 and 1680 spoken sentences. In this work, 6 noise types are artifically added by the authors at SNR similar to those of the AURORA2 test sets. Different instances of the same noise type are used for training and testing sets such that no instance of noise is ever repeated. Furthermore, in this work the test set is split into validation and test sets using a 1/3, 2/3 split. The purpose of the validation split is to find the optimal hyperparameter $\alpha$ shown in Figure \ref{fig:proposed_method}, while the test split is used to obtain the rest of the results in this work. The noise types used for training and testing are the same, meaning the noise types will be known under testing. The noise types \textit{babble, bus, caf} and \textit{pedestrian} are from the CHiME3 dataset \cite{CHIME_37} while \textit{babble} and \textit{speech shaped noise} are generated by the authors of \cite{TIMIT_noisetypes}. The noise is artifically added as described in section 2.A of \cite{adding_noise}. The VAD labels for the TIMIT database is generated using the wrd-formatted files in the database which states, at which time stamps speech is present. These labels are shared along with the source code of this work which is publicly available on GitHub.

\section{Experimental setup and results}

In order to evaluate the performance of the proposed method for adversarial training, the model is trained and tested on both the TIMIT database and the AURORA database. When evaluating Eq. (\ref{eq:context}) with the kernel sizes shown in Figure \ref{fig:proposed_method}, the context to generate a VAD label spans 398 ms to both sides. This means that to generate the first/last VAD label of each file, 398 ms of context is missing. In combination with the short duration of the files of AURORA2 (typically 0.8-2 seconds) and TIMIT (typically 2-4 seconds) leads to that a large part of the VAD outputs of each file will be generated without sufficient context. Additionally, the files in these data sets are all following the same structure. That is a short duration of silence in the beginning and the end, while the middle part contains speech. This leads to that the VAD algorithm learns to recognize this structure based on the missing context in the beginning and the end. To deal with this problem, during training and testing 10 files of the same noise type and SNR level are randomly chosen and concatenated leading to longer inputs to the VAD algorithm and therefore a smaller part of the VAD outputs will be generated based on insufficient context. The reason for using 10 files is that the computer on which the model is trained runs into memory problems when using more files. This forward CNN calculation step is performed three times before each backward step leading to a mini-batch size of 30 audio files. For each three forward steps a single backward step is performed using the RMSprop optimiser. The model is trained over 30 epochs, the learning rate is initialised as $0.01$ and the learning rate is multiplied by $0.7$ after each epoch.

\rowcolors{2}{white}{gray!25}
\begin{center}
\begin{table}[!htp]
\centering
\caption{AUC values on the validation sets using different values of $\alpha$}
\label{table:alphas}
\resizebox{0.45\textwidth}{!}{%
\begin{tabular}{|c || c | c  c  c  c  c |} 
 \hline
 $\alpha$ & \textbf{0} & \textbf{0.01} & \textbf{0.1} & \textbf{1}  & \textbf{10} & \textbf{100} \\ 
 \hline\hline
  \textbf{AURORA2 A} &93.90 & 94.15 & \textbf{95.18} & 94.74 & 94.30 & 94.24  \\ 
 \hline
  \textbf{AURORA2 B} & 91.15 & 91.16 & 92.49 & \textbf{92.69} & 91.13 & 90.38  \\ 
 \hline
  \textbf{TIMIT} & 88.78 & 89.98 & \textbf{90.32} & 89.3 & 88.91 & 87.94\\ 
 \hline
\end{tabular} }
\vspace{-10mm}
\end{table}
\end{center}

\subsection{Adversarial multi-task learning}
First the optimum value of the scalar $\alpha$ is found experimentally on both data sets by using validation data. Given that the AURORA2 is labelled as 73\% speech and TIMIT is labelled as 85\% speech, the results will be given by calculating the Area Under Curve (AUC) of their respective Receiver Operating Characteristics (ROC) curves while the accuracy will be disregarded as it can be misleading. For finding the optimum value of $\alpha$ the average AUC over the noise types of the validation split at each SNR level is calculated.
In each experiment the model is initialised using the same set of parameters to remove the randomness that can potentially be introduced by different initialisations. The Leaky ReLU layers are initialised using He \cite{He_ini} while the parameters of the sigmoid layers are initialised using Xavier \cite{Xavier_ini}. The Leaky ReLU layers use a slope coefficient of 0.01. 
The average AUC at $\alpha = 10^n, n \in [-2, -1, 0, 1, 2]$ on the validation sets are presented in Table \ref{table:alphas}. It is seen that the performance of the VAD is increased by a wide range of values of $\alpha$ and in general the addition of a discriminative network for adversarial multi-task learning outperforms the baseline model in terms of AUC.
In the case of all 3 test sets, and thereby also to the model both known and unknown noise types, it is found that a wide range of values of $\alpha$ results in an increase in performance. In two of three cases a value of $0.1$ is found to be optimal, thus this will be the value used for further experiments in this work.

The performance of the model using an $\alpha$ value of $0.1$, which was found optimal on the validation splits, is further investigated using the test splits of each data set. The results are seen in Table \ref{tab:different_alpha}. Once again the models are trained from the same initial values using the same simulation settings as described earlier and it is clearly seen that while the performance at high SNR levels is approximately the same with and without adversarial multi-task learning, at lower SNR levels the models trained using adversarial multi-task learning performs better and proves to be more noise-robust. This is the case both when it comes to noise types known to the model (TIMIT and AURORA2 test set A) and noise types unknown to the model (AURORA2 test set B).

\rowcolors{2}{white}{gray!25}
\begin{center}
\begin{table}[!htp]
\centering
\caption{AUC values on the test sets of AURORA2 and TIMIT with \textbf{(W)} and without \textbf{(W/O)} adversarial multi-task learning. When adversarial multi task learning is used, an $\alpha$ value of 0.1 is used}
\label{tab:different_alpha}
\resizebox{0.47\textwidth}{!}{%
\begin{tabular}{|c | c c c c c c c|c |} 
 \hline
  & Clean & 20 dB & 15 dB & 10 dB & 5 dB & 0 dB & -5 dB & \textbf{Mean} \\ 
  \hline\hline
  \textbf{AURORA2 A - W/O} & \textbf{97.96} & \textbf{97.95} & 97.85 & 97.62 & 96.59 & 92.12 & 78.79 & \textit{94.08}\\ 
 \hline
  \textbf{AURORA2 A - W} & 97.91 & 97.90 & \textbf{97.88} & \textbf{97.66} & \textbf{96.94} & \textbf{93.44} & \textbf{84.49} & \textit{\textbf{95.18}}\\ 
 \hline\hline
  \textbf{AURORA2 B - W/O} & \textbf{97.96} & \textbf{97.96} & 97.10 & 95.05 & 89.56 & 79.01 & 63.79 & \textit{88.64}\\ 
 \hline
  \textbf{AURORA2 B - W} & 97.91 & 97.90 & \textbf{97.54} & \textbf{96.54} & \textbf{94.20} & \textbf{89.11} & \textbf{74.25} & \textit{\textbf{92.49}}\\ 
   \hline\hline
  \textbf{TIMIT - W/O} & 95.44 & 95.41 & 94.63 & 92.64 & 88.60 & 82.67 & 71.96 & \textit{88.76}\\ 
 \hline
  \textbf{TIMIT - W} & \textbf{95.62} & \textbf{95.49 }& \textbf{94.92} & \textbf{93.91} & \textbf{90.75} & \textbf{84.94} & \textbf{74.26} & \textit{\textbf{89.99}}\\
 \hline
\end{tabular} }
\vspace{-6mm}
\end{table}
\end{center}

\subsection{Algorithmic delay}
The second part of this work is to evaluate the performance of the proposed method at lower algorithmic delays. The model is trained with the optimum value of $\alpha=0.1$ while the kernel sizes of the decoder block is reduced.
The algorithmic delay is calculated as a function of the sampling frequency and the kernel sizes as in Eq. (\ref{eq:context}). The majority of the algorithmic delay is introduced by the DB, thus only these kernel sizes will be varied. The kernel sizes and their corresponding algorithmic delays as used in the experiments are presented in Table \ref{table:context_DB}. The AUC is calculated as the average over the 7 SNR levels and the noise types of each test set.

\begin{figure}[!htp]
    \centering
    \includegraphics[width=0.45\textwidth]{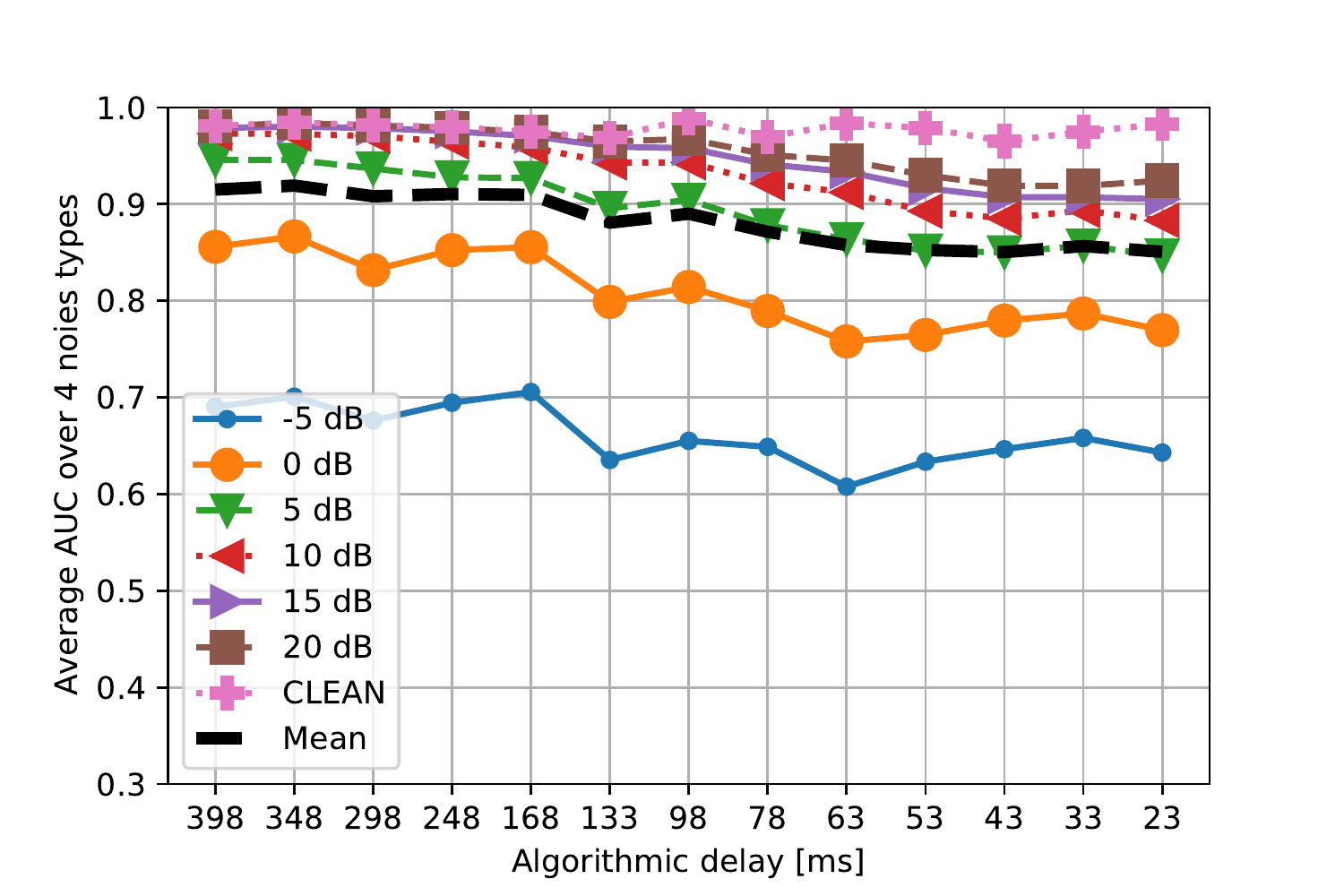}
    \caption{Average AUC at decreasing algorithmic delays on the (to the model) unknown noise types of AURORA2 test set B}
    \label{fig:performance_algDelay}
    \vspace{-5mm}
\end{figure}

It is seen that the performance of the VAD  decreases as the algorithmic delay is lowered, however this performance decrease is not drastic. It is in particular interesting to note that even when completely disregarding the decoder block with an algorithmic delay of 23 ms the VAD still performs well. When decreasing the algorithmic delay from 398 ms to 23 ms, only a performance decrease of 7\% AUC is seen. The performance of different algorithmic delays at each SNR level for AURORA2 test set B is presented in Figure \ref{fig:performance_algDelay}. In particular the performance at clean speech seems to be unaffected by the low algorithmic delay.
\vspace{-2mm}

\rowcolors{2}{white}{gray!25}
\begin{center}
\begin{table}[!htp]
\centering
\caption{VAD performance in terms of AUC at different kernel sizes and algorithmic delays tested on AURORA2 test sets A and B}
\label{table:context_DB}
\resizebox{0.42\textwidth}{!}{%
\begin{tabular}{|c c c | c | c | c|}  
 \hline
 \textbf{DB1} & \textbf{DB2} & \textbf{DB3} & \textbf{AD [ms]} & \textbf{AURORA2 B} & \textbf{AURORA2 A}\\ 
 \hline\hline
  55 & 15 & 5 & \textit{398}  & 92.11 & 94.44\\ 
 \hline
 45 & 15 & 5 & \textit{348}  &  91.91 & 94.38\\ 
 \hline
   35 & 15 & 5 & \textit{298} & 90.81 & 93.44\\ 
  \hline
  25 & 15 & 5 & \textit{248} & 91.04 & 93.32\\ 
 \hline
 15 & 10 & 5 & \textit{173} & 90.95 & 93.01\\
 \hline
  10 & 7 & 5 & \textit{133} & 88.09 & 90.84\\ 
 \hline 
  7 & 5 & 5 & \textit{98} & 89.01 & 91.85\\ 
 \hline
  5 & 3 & 3 & \textit{78} & 87.14 & 89.78\\
 \hline
  3 & 3 & 2 & \textit{63} & 85.77 & 90.00\\ 
 \hline
 2 & 2 & 2 & \textit{53} & 85.26 & 87.56\\
 \hline
 2 & 2 & 0 & \textit{43} & 85.03 & 87.40\\ 
 \hline
2 & 0 & 0 & \textit{33} & 85.66 & 87.90\\
 \hline
 0 & 0 & 0 & \textit{23} & 85.07 & 87.27\\ 
 \hline
\end{tabular} }
\vspace{-12mm}
\end{table}
\end{center}

\section{Conclusions}
In this paper we proposed a novel approach of training a supervised VAD using adversarial multi-task learning, where the model is trained friendly to the VAD labels and adversarially to the noise types aiming to make the VAD more invariant to noise. This is done by introducing an additional sub-network which aims to classify the noise types to the model. The VAD is then trained adversarially to these. It is found that the adversarial multi-task training is capable of increasing the VAD performance especially in more noisy environments, and it is shown to increase performance both when presented to unknown and already known noise types. At lower SNR levels the performance is boosted for both AURORA2 sets A and B. On the TIMIT set an increase is also observed, however less significant. This multi-task learning is only used when training the model and disregarded under testing, meaning the proposed method is cost-less once training has finished.

Furthermore it was investigated if this method can be useful in a low-latency application. This was done by reducing the kernel sizes of the DB resulting in lower algorithmic delays. It was found that even at an algorithmic delay as low as 23 ms, at which point the DB is completely disregarded, the performance of the method was still good. When decreasing the algorithmic delay from 398 ms to 23 ms the performance is only reducd by 7\% AUC.


\newpage

\bibliographystyle{IEEEtran}

\bibliography{mybib}


\end{document}